\documentclass[journal]{IEEEtran}
\usepackage{amsthm}
\usepackage{amsmath}
\usepackage{amssymb}
\usepackage{subfigure}
\usepackage{epic}
\newtheorem{theorem}{Theorem}
\newtheorem{lemma}{Lemma}
\newtheorem{corollary}{Corollary}
\newcommand{\Nb}{{\mathbb{N}}}
\newcommand{\Zb}{{\mathbb{Z}}}
\newcommand{\Qb}{{\mathbb{Q}}}

\newcommand{\B}{{\cal B}}

\newcommand{\C}{{\cal C}}

\theoremstyle{remark}
\newtheorem{remark}{Remark}

\newcommand{\bra}[1]{\langle #1\rangle}
\begin{document}

\title{Computational limits to nonparametric estimation for ergodic processes}
\author{Hayato Takahashi,~\IEEEmembership{Member,~IEEE}
\thanks{The author is with the Institute of Statistical Mathematics,
10-3 Midori-cho, Tachikawa, Tokyo 190-8562, Japan.
 e-mail: hayato.takahashi@ieee.org.}}
 \markboth{}{}
 \maketitle
\begin{abstract}
A new negative result for nonparametric distribution estimation of binary ergodic processes is shown. 
The problem of estimation of distribution with any degree of accuracy is studied.
Then it is shown that  for any countable class of estimators there is a zero-entropy binary ergodic process that is inconsistent with the class of estimators.
Our result is  different from other negative results for universal forecasting scheme of ergodic processes. 
\end{abstract}
\IEEEpeerreviewmaketitle

\section{Introduction}\label{sec-intro}
Let \(X_1, X_2,\ldots\) be a binary-valued ergodic process and \(P\) be its distribution.
In this paper we study nonparametric estimation of binary-valued ergodic processes with any degree of accuracy. 
Let \(S\) and \(\Omega\) be the set of finite binary strings and the set of infinite binary sequences, respectively. 
Let \(\Delta(x):=\{x\omega | \omega\in\Omega\}\),
where \(x\omega\) is the concatenation of \(x\in S\) and \(\omega\), and write \(P(x)=P(\Delta(x))\).
For \(x\in S\), \(|x|\) is the length of \(x\).
Let  \(\Nb\), \(\Zb\), and \(\Qb\) be the set of natural numbers, the set of integers, and the set of rational numbers, respectively. 
From ergodic theorem, there is a function \(r\) such that  for \(x\in S\), \(n,k\in\Nb\),
\begin{equation}\label{rate-func}
\begin{gathered}
 P(\cup\{\Delta(y)\mid |P(x)- \frac{1}{|y|}\sum_{i=1}^{|y|-|x|+1} I _{y_i^{i+|x|-1}=x}|\geq 1/k, \\
\hspace{4.5cm} |y|=n\})<r(n,k,x), \\
 \forall x,k\ \lim_n r(n,k,x)=0,
\end{gathered}
\end{equation}
where   \(I\) is the indicator function and \(y_i^j=y_iy_{i+1}\cdots y_j\) for \(y=y_1\cdots y_n, i\leq j\leq n\).
\(r\) is called convergence rate.
If \(r\) is given, we know how much sample size is necessary to estimate the distribution with prescribed accuracy. 
However it is known that there is no universal convergence rate for ergodic theorem. 
If \(r\) is not known, ergodic theorem does not help to estimate the distribution with prescribed accuracy. 
Here a natural question arise: for any binary-valued ergodic process, is it always possible to estimate the distribution with any degree of accuracy with positive probability? 
We show that this problem has a negative answer, i.e.,  for any countable class of estimators there is a zero-entropy binary ergodic process that is not estimated from this class of estimators with positive probability. 
In particular, since the set of computable functions is countable, we see that there is a zero-entropy binary ergodic process that is inconsistent with computable estimators. 
Our result is not derived from other negative results for universal forecasting scheme of ergodic processes, see Remark~\ref{rem-forecasting}.

Let \(x\sqsubseteq y\) if \(x\) is a prefix of \(y\).
\(f\) is called estimator if \(\exists D_f\subseteq S\times \Nb\times S \ f:D_f\to\Qb\) and
\begin{equation}\label{f-condition}
\begin{aligned}
&f(x,k,y)\mbox{ is defined, i.e.,  }(x,k,y)\in D_f\\
&\Rightarrow\forall z\sqsupseteq y\ f(x,k,z)=f(x,k,y).
\end{aligned}
\end{equation}
For \(\omega\in\Omega\), let \(f(x,k,\omega):=f(x,k,y)\) if \(f(x,k,y)\) is defined and \(y\sqsubset\omega\).
We say that \(f\) estimates \(P\) if
\begin{equation}\label{f-estimate}
\begin{aligned}
P(\omega \mid \forall x,k\ & f(x,k,\omega)\mbox{ is defined and }\\
&|P(x)-f(x,k,\omega)|<\frac{1}{k})>0.
\end{aligned}
\end{equation}
Here  \(\omega\) is a sample sequence and the minimum length of \(y\sqsubset\omega\) for which \(f(x,k,y)\) is defined is a stopping time.

In this paper, we construct an ergodic process that is not estimated from any given countable set of estimators:
\begin{theorem}\label{main-th-1}
\begin{align*}
&\forall F:\text{countable set of estimators~}\\
&\exists P\mbox{ ergodic and zero entropy }\forall f\in F\\
&P(\omega \mid \forall x,k\ f(x,k,\omega)\mbox{ is defined and }\\
&\hspace{2cm}|P(x)-f(x,k,\omega)|<\frac{1}{k})=0.
\end{align*}
\end{theorem}

We say that \(P\) is {\it effectively estimated} if there is a partial computable \(f\) that satisfies (\ref{f-condition}) and (\ref{f-estimate}).
Since the set of partial computable estimators is countable,  we have
\begin{corollary}\label{col-main-1}
There is a zero entropy ergodic process that is not effectively estimated. 
\end{corollary}

If \(r\) in (\ref{rate-func}) is computable then it is easy to see that  \(P\) is effectively estimated.\footnote{
More precisely if \(r\) is upper semi-computable (approximated from above by some algorithm), \(P\) is effectively estimated.}
For example, i.i.d.~processes of finite alphabet are effectively estimated, see Leeuw et~al.~\cite{Leeuw56}.

As stated above, a difficulty of effective estimation of ergodic processes comes from that  there is no universal convergence rate for ergodic theorem. 
In Shields pp.171 \cite{shields96}, it is shown that for any given decreasing function \(r\), there is an ergodic process that satisfies  
\begin{equation}\label{counter-example}
\exists N\forall n\geq N\ P(|P(1)-\sum_{i=1}^{n}I_{X_i=1}/n|\geq1/2)>r(n).
\end{equation}
In particular if \(r\) is chosen such that \(r\) decreases to 0 asymptotically slower than any computable function 
then \(r\) is not computable. 
In V'yugin \cite{vyugin98}, a binary-valued computable stationary process with incomputable convergence rate is shown.

It is possible that an ergodic process is effectively estimated even if the convergence rate is not computable (nor upper semi-computable).
\begin{theorem}\label{main-th-2}
For any decreasing \(r\), there is a zero entropy ergodic process that is effectively estimated and satisfies (\ref{counter-example}).
\end{theorem}

\begin{remark}\label{rem-forecasting}
In Cover \cite{cover70}, two problems about prediction of ergodic processes are posed. 
Problem 1 : Is there a universal scheme \(f\) such that 
\(\lim_{n\to\infty} | f(X_0^{n-1})-P(X_n|X_0^{n-1})|\to 0\), a.s.  for all binary-valued ergodic \(P\)? Problem 2 : Is there a universal scheme \(f\) such that 
\(\lim_{n\to\infty} | f(X_n^{-1})-P(X_0|X_{-\infty}^{-1})|\to 0\), a.s.  for all binary-valued ergodic \(P\)?
Problem 2  was affirmatively solved by Ornstein \cite{{ornstein78},{weiss00}}.
Problem 1 has a negative answer as follows (Bailey, Ryabko, see \cite{{bailey},{ryabko88},{gyorfi-etal}}):
For any \(f\) there is a binary-valued ergodic process \(X_1, X_2,\ldots\) such that
\begin{equation}\label{bailey-ryabko}
P(\limsup_{n\to\infty} | f(X_0^{n-1})-P(X_n |X_0^{n-1})|>0)>0.
\end{equation}

It is not difficult to see that the above result is extended to a countable class \(\{f_1,f_2,\ldots\}\), i.e., for any \(\{f_1,f_2,\ldots\}\) there is an ergodic process such that    (\ref{bailey-ryabko}) holds for all \(f_1,f_2,\ldots\).
However this result does not imply Theorem~\ref{main-th-1}.
In fact, there is a finite-valued ergodic process that is effectively estimated but satisfies (\ref{bailey-ryabko}), see below. 
Roughly speaking, one of the difference between these problems is that in Problem 1 we have to estimate \(P(X_n |X_0^{n-1})\) from \(X_0^{n-1}\),
however in our estimation scheme, sample size is a stopping time and we can use a sufficiently large sample \(X_0^m, m>n\) to estimate \(P(X_0^n)\).

In Ryabko \cite{ryabko88}, the process for (\ref{bailey-ryabko}) is constructed as follows:
First consider an ergodic Markov process \(Y_1, Y_2,\ldots\) on a countable state \(0, 1, 2, \ldots\) and \(P_{i, i+1}:=1/2, P_{i, 0}:=1/2\) for \(i= 0, 1,\ldots\), where \(P_{i,j}\) is the transition probability from \(i\) to \(j\).
The process \(X_i\in\{0,1,2\}, i\in\Nb\) is defined by \(X_i=0\) if \(Y_i=0\) and \(P(X_i=1)=p_j, P(X_i=2)=1-p_j\) if \(Y_i=j\geq 1\).
Then  \(\{X_i\}\) is ergodic.
In particular, for any \(\{f_1,f_2,\ldots\}\),  we can choose \(\{p_1,p_2,\ldots\}\) such that (\ref{bailey-ryabko}) holds for all \(f_i, i\in\Nb\).
However \(\{X_i\}\) is effectively estimated for any \(\{p_i\}\) as follows.
Let 
\[
I_j=\{i\mid X_i=0\text{ and } X_k\ne 0\text{ for }i<k\leq i+j\}.
\]
From the construction, we have
\begin{align*}
X_i=0\text{ and }X_k\ne 0\text{ for }i<k\leq i+j \Leftrightarrow  Y_{i+j}=j.
\end{align*}
Since the above event has a positive probability, \(I_j\) is an infinite set with probability one. 
Since \(Y_{i+j}=j\) for \(i\in I_j\),  \(\{X_{i+j}\}_{i\in I_j}\) are i.i.d.~random variables with \(P(X_{i+j}=1)=p_j, P(X_{i+j}=2)=1-p_j\).
Thus we can estimate \(p_j\) with any degree of accuracy for all \(j\).
Since the process \(\{X_i\}\) is determined from \(\{p_i\}\), it is effectively estimated. 
\end{remark}

\section{Cutting and Stacking}
We construct  ergodic processes in Theorem~\ref{main-th-1} and \ref{main-th-2} by cutting and stacking method. 
The basic idea of our constructions are similar to  that of (\ref{counter-example}).
In this section, we briefly introduce some notions about cutting and stacking, which we need in the proof. 

Let \(X:=[0,1]\) and consider Lebesgue measure \(\lambda\) on \((X,\B)\), where \(\B\) is the Borel \(\sigma\)-field. 
We construct an ergodic transformation \(T\) on \((X,\B,\lambda)\).
Let \(\C:=(L_1, L_2,\ldots, L_n)\) be an ordered set of mutually disjoint intervals of equal length.
\(\C\) is called column. 
\(w(\C):=\lambda(L_1)\), \(h(\C):=n\), and  \(S(\C):=\cup_i L_i\) are  called width, height, and support of \(\C\), respectively. 
Two columns are called disjoint if their support are disjoint. 
For two  disjoint columns \(\C:=(L_1, L_2,\ldots, L_n)\) and \(\C':=(L'_1, L'_2,\ldots, L'_m)\) of the same width,
let \(\C\ast \C' :=(L_1,\ldots,L_n,L'_1,\ldots,L'_m)\).
For a given column \(\C=\{L_i\}_{1\leq i\leq n}\), two disjoint columns \(\C_L:=\{L^1_i\}_{1\leq i\leq n}\) and \(\C_R:=\{L^2_i\}_{1\leq i\leq n}\) are called partition of \(\C\)
if \(L_i=L^1_i\cup L^2_i\) and \(L^1_i\cap L^2_i=\emptyset\) for \(1\leq i\leq n\) and \(w(\C_R)=w(\C_L)=\frac{1}{2}w(\C)\).
In order to specify the partition, we require that the left-endpoint of \(L^1_i\) is less than that of \(L^2_i\).
Let \(\C\ast\C:=\C_L\ast\C_R\), where \(\C_L\) and \(\C_R\) are partition of \(\C\), see Fig.~\ref{fig-1}.
Let \(\C(0):=\C\) and \(\C(n+1):=\C(n)\ast\C(n)\) for \(n\geq 0\).
We have \(w(\C(n+1))=\frac{1}{2}w(\C(n))=2^{-(n+1)}w(\C)\) and \(h(\C(n+1))=2h(\C(n))=2^{n+1}h(\C)\).

A transformation \(T\) is defined on a column  \(\C:=(L_1, \ldots,L_n)\) by 1) \(T(L_i)=L_{i+1}\) and \(T(a_i+y)=a_{i+1}+y\), where \(0\leq y\leq w(\C)\), and \(a_i\) is the left-endpoint of \(L_i\)  for \(1\leq i\leq n-1\), 
and 2) \(T\) is not defined on \(L_n\).
Then \(T\) is a  measure preserving transformation defined on intervals of \(\C\) except for \(L_n\).
Similarly, \(T^{-1}\) is  defined on \(\C\) except for \(L_1\).
Note that \(T\) (and \(T^{-1}\)) defined by \(\C\ast\C\) extends \(T\)  (and \(T^{-1}\)) defined by \(\C\), respectively. 

We say that a sequence of columns \(\C_1,\C_2,\ldots\) is {\it extending} if  \(S(\C_n)\subseteq S(\C_{n+1})\) and \(T\) defined by \(\C_{n+1}\) extends \(T\) defined by \(\C_n\) for all \(n\).
Suppose that there is an extending sequence of columns \(\C_1,\C_2,\ldots\) such that  \(\lim_n w(\C_n)=0\) and \(\lambda(\cup_n S(\C_n))=1\).
Then we see that an invertible measure-preserving transformation \(T: X\to X\) is uniquely defined except for a null set.
 \(T\) is ergodic as follows:
Suppose that \(A\subseteq X\) is a nontrivial invariant set, i.e., \(T(A)=A\) and \(0<\lambda(A)<1\).
Since \(\lim_n w(\C_n)=0\) and \(\lambda(\cup_n S(\C_n))=1\), from Lebesgue density theorem, there are \(n\) and \(L_i,L_j\in\C_n\) such that 
\(1/2<\lambda(A\cap L_i)/\lambda(L_i), 1/2<\lambda(A^c\cap L_j)/\lambda(L_j)\).
Then \(\lambda(A\cap A^c)=\lambda(T^{j-i}(A)\cap A^c)\geq \lambda(T^{j-i}(A\cap L_i)\cap A^c\cap L_j)>0\), which is a contradiction.

Let \(X^0\) and \(X^1\) be measurable sets of \(X\) such that \(X^0\cup X^1=X\) and \(X^0\cap X^1=\emptyset\).
For \(\xi\in X\), let \(\phi(\xi)=\cdots\xi(-1)\xi(0)\xi(1)\cdots\in\{0,1\}^\Zb\), where
\(\xi(i)=0\) if \(T^i(\xi)\in X^0\) and \(1\) else for all \(i\in\Zb\).
Let \(P:=\lambda\circ\phi^{-1}\).
If \(T\) is an invertible ergodic transformation, \(P\) is an invertible ergodic process on \(\{0,1\}^\Zb\) and is called \((T, X^0,X^1)\) process. 
We say that a column \((L_1,\ldots,L_n)\) is {\it compatible} with \((X^0,X^1)\) if \(\forall1\leq i\leq n, L_i\subseteq X^0\mbox{ or }L_i\subseteq X^1\), and 
in that case let \(s(L_i):=0\) if \(L_i\subseteq X^0\) and 1 else for \(1\leq i\leq n\), and \(s(L_1,\ldots,L_n):=s(L_1)\cdots s(L_n)\).

\setlength{\unitlength}{1mm}
\newcommand{\columnA}{
\begin{picture}(30,30)
\put(5,5){\line(1,0){20}}
\put(0,5){\(L_1\)}
\put(15,6){\vector(0,1){5}}
\put(5,13){\line(1,0){20}}
\put(0,13){\(L_2\)}
\put(15,17){\(\cdot\)}
\put(15,20){\(\cdot\)}
\put(0,25){\(L_n\)}
\put(5,25){\line(1,0){20}}
\end{picture}}
\newcommand{\columnB}{
\begin{picture}(30,30)
\put(5,5){\line(1,0){20}}
\put(0,5){\(L_1\)}
\put(10,6){\vector(0,1){5}}
\dashline{1}(15,3)(15,27)
\put(5,13){\line(1,0){20}}
\put(0,13){\(L_2\)}
\put(10,17){\(\cdot\)}
\put(10,20){\(\cdot\)}
\put(20,17){\(\cdot\)}
\put(20,20){\(\cdot\)}
\put(0,25){\(L_n\)}
\put(5,25){\line(1,0){20}}
\put(11,23){\vector(1,-2){8}}
\put(20,6){\vector(0,1){5}}
\end{picture}}
\def\goodgap{
\hspace{\subfigtopskip}
\hspace{\subfigbottomskip}
}
\begin{figure}[hbt]
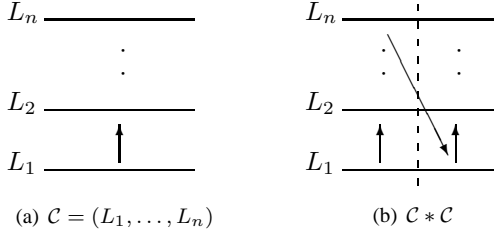

\begin{center}
\begin{tabular}[t]{c}
\subfigure[\(\C=(L_1,\ldots,L_n)\)]{\columnA}
\goodgap
\subfigure[\(\C\ast\C\)]{\columnB}
\end{tabular}
\end{center}
\caption{Cutting and stacking}
\label{fig-1}
\end{figure}

\section{Proof of Theorem~\ref{main-th-1}}
Let \(F:=\{f_1,f_2,\ldots\}\) be a countable set of estimators. 
Consider the following three statements:
\begin{equation}\label{A:1}
\begin{aligned}
&\forall P\text{ ergodic on }\Omega~\exists f\in F\\
& P(\omega \mid \forall x,k\ |P(x)-f(x,k,\omega)|<\frac{1}{k})>0,
\end{aligned}
\end{equation}
\begin{equation}\label{A:2}
\begin{aligned}
&\forall P\text{ ergodic on }\Omega~\exists f\in F\\
& P(\omega \mid \forall m,k\ |P(0^m)-f(0^m,k,\omega)|<\frac{1}{k})>0,
\end{aligned}
\end{equation}
\begin{equation}\label{A:3}
\forall P\text{ ergodic on }\Omega~\exists \hat{f}\in \hat{F}\ P(\omega \mid \hat{f}(\omega)=R)>0,
\end{equation}
where \(0^m\) is the \(m\)-times concatenation of \(0\)'s,
\begin{gather*}
R:=\{(n,m)\mid P(0^m)<2^{-(n+2)}\},\\
\hat{f}:=\{(n,m,y) \mid \exists k\ f(0^m,k,y)+\frac{1}{k}<2^{-(n+2)}\},\\
 \hat{F}:=\{\hat{f} | f\in F\},\\
 \hat{f}(x):=\{(n,m)\mid (n,m,y)\in\hat{f}, y\sqsubseteq x\},\ \hat{f}(\omega):=\cup_{x\sqsubset\omega}\hat{f}(x).
\end{gather*}
Then we have (\ref{A:1})\(\Rightarrow\)(\ref{A:2})\(\Rightarrow\)(\ref{A:3}), where (\ref{A:2})\(\Rightarrow\)(\ref{A:3}) follows from 
\[
\forall m,k\ |P(0^m)-f(0^m,k,\omega)|<\frac{1}{k}\Rightarrow \hat{f}(\omega)=R.
\]
Therefore in order to show Theorem~\ref{main-th-1}, it is sufficient to negate (\ref{A:3}) (Lemma~\ref{th-main-lemma}).

\begin{lemma}\label{th-main-lemma}
For any countable \(F\), there is a zero entropy ergodic \(P\) such that \\
\(\forall \hat{f}\in\hat{F},\ P(\omega \mid \hat{f}(\omega)=R)=0\).
\end{lemma}
Proof)
Let \(\hat{F}:=\{\hat{f}_1,\hat{f}_2,\ldots\}\).
We construct an ergodic process inductively by cutting and stacking method such that if there are \(a, x, e\) such that \(a\in \hat{f}_e(x)\) at some stage then
 the process is made to falsify \(\hat{f}_e\), i.e., \(a\notin R\).

Let \(X^0:=[0,1/2)\) and \(X^1:=[1/2,1]\).
For \(n\geq 1\), let \(A_n:=(2^{-(n+1)},2^{-n}]\). 
We construct inductively an extending sequence of columns \(\C_0,\C_1,\ldots\), which are compatible with \((X^0,X^1)\), \(\lim_n w(\C_n)=0\), and \(\cup_n S(\C_n)=\cup_{i\in J} A_i\cup X^1\),
where \(J\)  is defined simultaneously with columns.  

Stage \(0\): Let \(\C_0:=X^1\), \(G_0:=\emptyset\), and \(k_0:=1\).

Stage \(n\): Suppose that \(G_{n-1}\) is defined and \(\C_0,\ldots,\C_{n-1}\) are extending and compatible with \((X^0,X^1)\).
Let \(\C_{n-1}:=(L_1,\ldots,L_{h_{n-1}})\) and suppose that \(w(\C_{n-1})=2^{-k_{n-1}}\) for \(k_{n-1}\in\Nb\).
Let 
\begin{equation}\label{Fn-def}
\begin{aligned}
F_n:=\{(\bra{e,i},m)\mid & 1\leq e\leq n, 1\leq i\leq h_{n-1}, \\
&(\bra{e,i},m)\in \hat{f}_e(s(L_i\cdots L_{h_{n-1}}))\},
\end{aligned}
\end{equation}
\[
G_n:=\{\bra{e,i} \mid \exists m\ (\bra{e,i},m)\in F_n\}\cap(G_{n-1})^c,
\]
where \(\bra{\cdot,\cdot}:\Nb\times\Nb\to\Nb\) is a bijection and \((G_{n-1})^c\) is the complement of \(G_{n-1}\).\\
If \(G_n=\emptyset\) then set \(k_n:=k_{n-1}+1\) and \(\C_n:=\C_{n-1}(1)\).\\
If \(G_n\ne\emptyset\) then let
\[m(e,i):=\min\{m \mid (\bra{e,i},m)\in F_n\}\mbox{ for }\bra{e,i}\in G_n,\mbox{ and }\]
\begin{equation}\label{kn}
\begin{aligned}
&k_n:=\\
&\hspace{0.3cm}\max\{k_{n-1}+1,\\
&\hspace{1.2cm}\min\{t\in\Nb \mid \forall \bra{e,i}\in G_n, 2^{t-\bra{e,i}-1}\geq 2m(e,i)\}\}.
\end{aligned}
\end{equation}
Since
\(w(\C_{n-1})=2^{-k_{n-1}}\) and 
\(w(A_{\bra{e,i}})=2^{-(\bra{e,i}+1)}\), we have 
\begin{equation}\label{width}
w(\C_{n-1}(k_n-k_{n-1}))=w(A_{\bra{e,i}}(k_n-\bra{e,i}-1))=2^{-k_n}.
\end{equation}
Define
\begin{equation}\label{def-cn}
\C_n:=\C_{n-1}(k_n-k_{n-1})\ast A_{n_1}(k_n-n_1-1)\ast\cdots\ast A_{n_t}(k_n-n_t-1),
\end{equation}
where \(G_n=\{n_1<n_2<\cdots<n_t\}\).

By induction, we have constructed an extending sequence of columns \(\C_0,\C_1,\ldots\), which are compatible with \((X^0,X^1)\) and \(w(\C_n)=2^{-k_n}\).
Let \(J:=\cup_n G_n\) then  \(\cup_n S(\C_n)=\cup_{i\in J} A_i\cup X^1\).

Let \(\Omega_1:=\cup_n S(\C_n)\).
Let \(T: \Omega_1\to\Omega_1\) be an  invertible measure preserving transformation  defined by \(\cup_n \C_n\) and  \( P\) be the \((T,  X^0, X^1)\) process, then \(P\) is ergodic.

Let
\[R:=\{ (n,m)\mid P(0^m)<2^{-(n+2)}\}.\]
Suppose that  
there is an \(e\) such that
\begin{equation}\label{R-estimate}
P(\omega \mid \hat{f}_e(\omega)=R)>0.
\end{equation}
Since \(k_{n-1}<k_n\) for all \(n\),  we have \(\lim_n h(\C_n)=\infty\). 
Since \(\forall n, s(\C_{n-1})\sqsubseteq s(\C_n)\),  we see that \(s(\C_1), s(\C_2),\ldots\) defines a unique sequence \(\alpha:=\alpha_1\alpha_2\cdots\in\Omega, \forall i, \alpha_i\in\{0,1\}\) in the limit,
i.e., \(\forall n, s(\C_n)\sqsubset\alpha\).
Since \(\Omega_1=\cup_n S(\C_n)\), we have
\begin{equation}\label{xi-alpha}
\begin{split}
&\xi\in\Omega_1\Leftrightarrow\exists n,i, 1\leq i\leq h_n, \ \xi\in L_i,\ \C_n=(L_1,\ldots,L_{h_n}),\\
&\exists i\ \xi(0)\cdots\xi(h_n-i)=\alpha_i\cdots\alpha_{h_n},
\end{split}
\end{equation}
where \(h_n=h(\C_n)\).
From (\ref{Fn-def}), (\ref{R-estimate}), and (\ref{xi-alpha}), there are \(i,n\in\Nb, 1\leq i\leq h_{n-1}\) such that 
\begin{equation}\label{hypo}
\hat{f}_e(\alpha_i\cdots\alpha_{h_{n-1}})\subseteq R\mbox{ and }\bra{e,i}\in G_n.
\end{equation}
From (\ref{kn}), we have 
\[
s(A_{\bra{e,i}}(k_n-\bra{e,i}-1))=0^{2^{k_n-\bra{e,i}-1}}\sqsupseteq 0^{2m(e,i)}.
\]
Let \((L_1,\ldots,L_h):=A_{\bra{e,i}}(k_n-\bra{e,i}-1)\). 
If \(\xi\in\cup_{1\leq j\leq h/2}L_j\) then \(\phi(\xi)\sqsupseteq 0^{m(e,i)}\).
Since \(\lambda(S(A_{\bra{e,i}}))=2^{-(\bra{e,i}+1)}\), 
we have 
\[
P(0^{m(e,i)})\geq\lambda(\cup_{1\leq j\leq h/2}L_j)= 2^{-(\bra{e,i}+2)}.
\]
Then  \((\bra{e,i}, m(e,i))\in \hat{f}_e(\alpha_i\alpha_{i+1}\cdots)\) and \((\bra{e,i}, m(e,i))\notin R\), which contradicts to (\ref{hypo}), see Fig.~\ref{fig-3}.
Thus we have \(\forall e, P(\omega \mid \hat{f}_e(\omega)=R)=0\).

Next we show that the entropy is zero.
From (\ref{width}), (\ref{def-cn}), and (\ref{xi-alpha}), for \(1\leq i\leq j\leq h_n\), we have
\(P(\alpha_i\cdots\alpha_j)\geq\lambda(w(\C_n))=2^{-k_n}\) and \(h_n\geq 2^{k_n-k_{n-1}}h_{n-1}\geq 2^{k_n-k_0}h_0=2^{k_n-1}\).
Since \(1/2\leq\lambda(S(\C_n))\), we have
\(\frac{1}{4}\leq\lambda (\cup_{1\leq i\leq h_n/2} L_i)\) and
\begin{equation}\label{entropy-alpha}
\forall n\ P( \frac{-\log_2 P(\omega_1\cdots\omega_{h_n/2})}{h_n/2}\leq k_n2^{-k_n+2})\geq 1/4.
\end{equation}
Suppose that the entropy of \(P\) is positive. 
Since \(\lim_n k_n=\infty\) and \(\lim_n h_n=\infty\), from Shannon-McMillan-Breiman theorem, 
\[
\forall\epsilon\exists N\forall n\geq N\ P(-\log_2 P(\omega_1\cdots\omega_n)/n> k_n2^{-k_n+2})>1-\epsilon,
\]
which contradicts to (\ref{entropy-alpha}), and the entropy of \(P\) is zero. 
\qed

\begin{figure}[hbt]
\begin{center}
\begin{gather*}
s(\C_{n-1}) =\alpha_1\cdots\alpha_i\cdots\alpha_h,\ \hat{f}_e(\alpha_i\cdots\alpha_h)\ni (\bra{e,i},m)\\
\Downarrow\\
s(\C_n)  =\overbrace{\alpha_1\cdots\alpha_h}^{s(\C_{n-1})} \cdots \overbrace{\alpha_1\cdots\alpha_h}^{s(\C_{n-1})}\ 0\cdot \overbrace{\cdot\cdot 0\ 0\cdot\cdot\cdot 0}^{\sqsupset 0^m}.
\end{gather*}
\end{center}
\caption{Example of construction.
For simplicity, suppose that \(s(\C_{n-1})=\alpha_1\cdots\alpha_h\),  \(G_n=\{\bra{e,i}\}\), and  \((\bra{e,i},m)\in\hat{f}_e(\alpha_i\cdots\alpha_h)\).
Then by stacking a long column of \(A_{\bra{e,i}}\), \(\hat{f}_e(\alpha_i\cdots\alpha_h)\) fails to guess \(R\). We choose \(a,b\) such that 
\(w(\C_{n-1}(a))=w(A_{\bra{e,i}}(b)),\ 2^b\geq 2m\) and let \(\C_n:=\C_{n-1}(a)\ast  A_{\bra{e,i}}(b)\).
Then by considering the trajectories starting from the first half levels of  \(A_{\bra{e,i}}(b)\), we have  \(P(0^m)\geq \lambda (S(A_{\bra{e,i}}))/2=2^{-(\bra{e,i}+2)}\).
 }
\label{fig-3}
\end{figure}

\section{Proof of Theorem~\ref{main-th-2}}
\subsection{Construction of a process for (\ref{counter-example})}\label{sub-sec-construction}
Here we summarize the construction of the process for (\ref{counter-example}), which we use in Theorem~\ref{main-th-2}.
(Actually independent cutting and stacking method is used in \cite{shields96}, however we need not it here.)

We construct an extending sequence of columns \(\C_n, n=0,1,2,\ldots\) from \(k_0=1<k_1<k_2<\cdots\in\Nb\) by induction. 
Let \(X:=[0,1],\ X^0=[0,1/2],\ X^1=(1/2,1], \C_0:=X^1\), and \(A_n:=(2^{-(n+1)},2^{-n}]\) for \(n=1,2,\ldots\).

Stage \(n\): Suppose that \(\C_{n-1}\) is defined and \(w(\C_{n-1})=2^{-k_{n-1}}\). 
Since \(w(\C_{n-1}(k_n-k_{n-1}))=w(A_n(k_n-(n+1)))=2^{-k_n}\),
define
\begin{equation}\label{construct-2}
\C_n:=\C_{n-1}(k_n-k_{n-1})\ast A_n(k_n-(n+1)).
\end{equation}
Then 
\begin{equation}\label{construct-2a}
\begin{aligned}
&w(\C_n)=2^{-k_n},\\
&S(\C_n)= \cup_{i=1}^n A_i\cup X^1\mbox{ and }\lambda(S(\C_n))=1-2^{-(n+1)},\\
&h(\C_n)=\lambda(S(\C_n))/w(\C_n)=2^{k_n}(1-2^{-(n+1)}).
\end{aligned}
\end{equation}
Since \(\lim_n w(\C_n)=0\) and \(\lambda(\cup_n \C_n)=1\), \(\cup_n \C_n\) defines an invertible ergodic process \(T\). 
Let \(P\) be the \((T, X^0, X^1)\) process. 
Let \(A_n(k_n-(n+1)))=(L_1,\ldots,L_h), h=2^{k_n-(n+1)}\).
Since \(A_n\subseteq X^0\), we have \(s(L_1,\ldots,L_h)=0^h\) and 
 if \(\xi\in \cup_{i=1}^{h'} L_i\) then \(\xi(0)\xi(1)\cdots\xi(h'-1)=0^{h'}\) for \(h'=h/2\).
Since \(\lambda(\cup_{i=1}^{h'} L_i)=2^{-(n+2)}\), we have
\[
2^{-(n+2)}< P(0^{h'})\leq P(|P(0)-\sum_{i=1}^{h'}I_{X_i=0}/h'|\geq1/2).
\]
Thus by choosing \(\{k_i\}\), we can construct an ergodic process with arbitrary slow convergence rate.

\subsection{Proof}
We show Theorem~\ref{main-th-2} for the ergodic process defined in Section~\ref{sub-sec-construction}.

In the following we write \(x^n\) as the \(n\)-times concatenation of \(x\in S\), e.g., \((01)^2=0101\).
From (\ref{construct-2}), we have
\begin{equation}\label{symbols}
s(\C_n)=(s(\C_{n-1}))^{2^{k_n-k_{n-1}}}0^{2^{k_n-(n+1)}}.
\end{equation}
For example, if \(k_0=1, k_1=2, k_2=3\) then
\[s(\C_0)=1,\ s(\C_1)=110,\ s(\C_2)=1101100.\]
From (\ref{construct-2}),  (\ref{construct-2a}), and  (\ref{symbols}), we see that 
\begin{equation}\label{p-1}
\begin{aligned}
&P(01^{2^{k_1-1}}0)=w(\C_1)=2^{-k_1},\\
&P(01^n0)=0\mbox{ if }n\ne 2^{k_1-1}\mbox{ and }n\ne 0,\\
&P(10^{f(n)}1)=w(\C_n)=2^{-k_n},\ f(n)=\sum_{i=1}^n2^{k_i-(i+1)},\\
&P(10^m1)=0\mbox{ if }\forall n\ m\ne f(n)\mbox{ and }m\ne 0.
\end{aligned}
\end{equation}
Let 
\begin{equation}\label{An}
B_0:=\C_0\mbox{ and }B_n:=\cup_{i=1}^{(2^{k_n-k_{n-1}}-1)h(\C_{n-1})}L_i,
\end{equation}
for \(\C_n=(L_1,\ldots, L_{h_n}),\ n\geq 1\).
From  (\ref{construct-2}) and Lemma~\ref{lemma-th-2} below, we have
\[\lambda(\cap_{i=0}^n B_i)=\lambda(B_0)\Pi_{i=1}^n (1-2^{-(k_i-k_{i-1})}),\]
see Fig.~\ref{fig-2}.
Assume that \(\forall i\ k_i-k_{i-1}\geq i\). Since \(\sum_{i=1}^\infty 2^{-(k_i-k_{i-1})}\leq 1\), we have
\[\lambda(\cap_{i=0}^\infty B_i)>0.\]

Let \(\xi\in\cap_{i=0}^\infty B_i\) and \(\phi(\xi)':=\xi(0)\xi(1)\cdots\in\Omega\).
Then \\

(\(\ast\))   
the first time that the pattern \(10^n1\) appears in \(\phi(\xi)'\) is less than that of \(10^m1\) if \(n<m\), \(P(10^n1)>0\), and \(P(10^m1)>0\).\\

We have (\(\ast\)) as follows: Let \(\xi\in\cap_{i=0}^\infty B_i\).
 Let \(B(x):=\{ n \mid 10^n1\mbox{ appear in }x\}\), where we write \(10^01=11\).
Since \(k_1-k_0\geq 1\), from  (\ref{construct-2}) and  (\ref{An}), we have \(11\sqsubset \phi(\xi)'\).
Let \(x_0:=11\) then  (\(\ast\)) trivially holds for \(x_0\) and \(B(x_0)=\{0\}\).
From (\ref{construct-2}) and  (\ref{An}), there are \(x_{n-1}\in S\) and \(k\geq 1\) such that   \(x_{n-1}y1\sqsubset\phi(\xi)'\) for \(y:=s(\C_{n-1})^k0^{2^{k_n-(n+1)}}\).
Suppose that  (\(\ast\)) holds for \(x_{n-1}\) and \(B(x_{n-1})=\{f(i) \mid 0\leq i\leq n-1\}\), where \(f(0)=0\).
Since \(B(s(\C_{n-1}))\subseteq B(x_{n-1})\) and the first bit of \(s(\C_{n-1})\) is \(1\),
we have \(B(x_{n-1})=B(z)\) for \(x_{n-1}\sqsubseteq z\sqsubseteq x_{n-1}y\) and \(B(x_{n-1}y1)=B(x_{n-1})\cup\{f(n)\}\).
Let \(x_n:=x_{n-1}y1\) then  (\(\ast\)) holds for \(x_n\) and \(B(x_n)=\{f(i) \mid 0\leq i\leq n\}\).
By induction,  (\(\ast\)) holds for \(\phi(\xi)'\), see Fig.~\ref{fig-2}.

\begin{figure}[hbt]
\begin{center}
\begin{gather*}
\C_n=\overbrace{\C_{n-1}^1\ast\cdots}^{B_n}\ast~\C_{n-1}^{2^{k_n-k_{n-1}}}\ast A(k_n-(n+1)),
\end{gather*}
\goodgap
{\footnotesize
\[
\setcounter{MaxMatrixCols}{20}
\begin{matrix}
L_1 & L_2 & L_3 & \cdots & L_7 & L_8 & L_9 & L_{10} & L_{11} & L_{12} & L_{13} & L_{14} \\
1 & 1 & 0 & \cdots & 1 & 1 & 0 & 1 & 1 & 0 & 0 & 0.\\
\end{matrix}
\]}
\end{center}
\caption{\(\C_{n-1}^i, 1\leq i\leq 2^{k_n-k_{n-1}}\) is a \(2^{k_n-k_{n-1}}\) partition of \(\C_{n-1}\).
Note that  \(s(\C_{n-1})\) ends with \(10^{f(n-1)}\) and does not contain the pattern  \(10^{f(n-1)}1\).
Since \(s(\C_{n-1})\) starts with 1, the trajectories starting from \(B_n\) contain the pattern \(10^{f(n-1)}1\).
For example, 
 let \(k_0=1, k_1=2, k_2=4\) then \(s(\C_0)=1, s(\C_1)=110, s(\C_2)=11011011011000\).
\(B_0\) is the union of \(L_i\) such that \(s(L_i)=1\).
 \(B_0\cap B_1=L_1\cup L_4\cup L_7\cup L_{10}\), and
 \(B_0\cap B_1\cap B_2=L_1\cup L_4\cup L_7\).
 The trajectories starting from \(B_0\cap B_1\) always contain the pattern \(11\) and those from \(B_0\cap B_1\cap B_2\) always contain patterns \(11\) and \(101\).
 }
\label{fig-2}
\end{figure}
Let \(K:=\{ (i,k_i) \mid i\in \Nb\}\).
Since \(10^n1\) and \(01^m1\) appear in \(\phi(\xi)'\) iff \(P(10^n1)>0\) and \(P(01^m1)>0\).
From  (\ref{p-1}) and (\(\ast)\), we can compute \(K\) from \(\phi(\xi)'\) if \(\xi\in\cap_i B_i\).
Thus there is a partial computable \(g\)  such that 
(i) \(g(x)\) is defined then \(g(x)=g(z)\) for \(x\sqsubseteq z\), (ii) \(g(\omega):=\cup_{x\sqsubset\omega} g(x)\) and (iii)
\begin{equation}\label{lem-B}
P\{\omega \mid g(\omega)=K\}\geq \lambda(\cap_{i=0}^\infty B_i)>0.
\end{equation}

Let
\[P_n(x):=\lambda(\cup\{L_i \mid \exists 1\leq i\leq j\leq h,\  x=s(L_i\ldots L_j)\}),\]
for \(\C_n=(L_1,\ldots,L_h)\).
We have \(P_n(x)\leq P_{n+1}(x)\) and \(P_n(x)\geq P_n(x0)+P_n(x1)\) for all \(n\) and \(x\).
Since (i) \(P_n(x)\) is computable from \(s(\C_n)\) and \(w(\C_n)=2^{-k_n}\) and (ii) \(s(\C_n)\) is computable from \(k_0,\ldots,k_n\),  we have that \(P_n(x)\) is computable from \(k_0,\ldots,k_n\).
Since \(\lambda(\cup_n S(\C_n))=1\), we have \(\lim_n P_n(x)=P(x)\).
Since \(P\) is a probability, we can compute \(P(x)\) with any given precision from \(K\).
Thus \(P\) is effectively estimated from \(\phi(\xi)', \xi\in\cap_i B_i\).

Finally we show that the entropy is zero. 
Since \(w(\C_n)=2^{-k_n}\) and \(h(\C_n)=2^{k_n}(1-2^{-(n+1)})\), we have
\[\lim_n \frac{-\log_2 P(s(\C_n))}{h(\C_n)}=0.\]
Since \(\lambda(\cup_n S(\C_n))=1\), from a similar argument for the previous theorem, we see that the entropy  is zero. 
\qed

\begin{lemma}\label{lemma-th-2}
Let \(\C:=(L_1,\ldots,L_h)\) and \((L'_1,\ldots,L'_{2^kh}):=\C(k)\).
Then for \(0\leq n\leq 2^k-1\) and \(J\subseteq\{1,\ldots,h\}\),
\[
\begin{split}
& \cup_{j\in J}L_j\cap \cup_{i=nh+1}^{(n+1)h}L'_i=\cup_{j\in J'}L'_j,\  J'=\{j+nh | j\in J\},\\
& \lambda(\cup_{j\in J'}L'_j)=2^{-k}\lambda(\cup_{j\in J}L_j).
\end{split}
\]
\end{lemma}
Proof)
Since \(\C(k)\) is a concatenation of \(2^k\) columns of the same width partition of \(\C\),
the lemma follows. 
\qed

\begin{center}
{\bf Acknowledgement}
\end{center}
The author thanks Prof.~Teturo Kamae (Matsuyama Univ.), Prof.~Benjamin Weiss (Hebrew Univ.), and anonymous referees for helpful discussions and valuable comments.

\end{document}